\documentclass[aps,twocolumn,amsmath,showkeys,amssymb,superscriptaddress,longbibliography,floatfix, pre]{revtex4-2}

\pdfoutput=1
\usepackage{bm}
\usepackage{mathrsfs}
\usepackage{graphicx}
\usepackage{natbib}
\usepackage{latexsym}
\usepackage{amsmath, amssymb, amsthm}
\usepackage{bbold}
\usepackage{mathtools}
\usepackage{xcolor}
\usepackage{comment}
\usepackage{wasysym}
\usepackage[T1]{fontenc}
\usepackage[bottom]{footmisc}
\usepackage{epigraph}
\usepackage{lipsum}
\usepackage{float}
\usepackage[normalem]{ulem}
\usepackage{appendix}
\usepackage{hyperref}
\usepackage{multirow} 
\usepackage{booktabs}
\usepackage{comment}
\usepackage{makecell}
\graphicspath{{images/}} 

\def\be{\begin{equation}}
\def\ee{\end{equation}}
\def\bea{\begin{eqnarray}}
\def\eea{\end{eqnarray}}

\newcommand\arrowpm{\overset{(\pm)}{\rightarrow}}
\newcommand{\avg}[1]{\langle{#1}\rangle}
\newcommand{\Avg}[1]{\left\langle{#1}\right\rangle}

\begin{document}

\title{Directed extended-range percolation}

\author{Wenbo Liu}
 \affiliation{School of Artificial Intelligence and Automation, Institute of Medical Equipment Science and Engineering, and State Key Laboratory of Digital Manufacturing Equipments and Technology, Huazhong University of Science and Technology, Wuhan, China}
\author{Yiwen Zeng}
 \affiliation{School of Artificial Intelligence and Automation, Institute of Medical Equipment Science and Engineering, and State Key Laboratory of Digital Manufacturing Equipments and Technology, Huazhong University of Science and Technology, Wuhan, China}
 \author{Xueming Liu}
  \email{xm\_liu@hust.edu.cn}
 \affiliation{School of Artificial Intelligence and Automation, Institute of Medical Equipment Science and Engineering, and State Key Laboratory of Digital Manufacturing Equipments and Technology, Huazhong University of Science and Technology, Wuhan, China}
 \author{Ginestra Bianconi}
  \email{ginestra.bianconi@gmail.com}
 \affiliation{School of Mathematical Sciences, Queen Mary University of London, London E1 4NS, UK}

\begin{abstract}
While for standard percolation directionality is known to increase the combinatorial complexity of percolation, here we show that when connectivity is ensured by paths of length $R\geq 2$,  network directionality, impeding backtracking,  can significantly reduce the complexity of percolation. To illustrate this finding, we introduce Directed Extended-Range Percolation (DERP), defined directed networks with non-reciprocal edges, motivated by applications in quantum communication. In this framework, message transmission is enabled between trusted nodes separated by a directed path of length at most $R$.  Using a message-passing approach, we show that directionality enables an exact determination of the percolation threshold and the anomalous critical indices on locally tree-like structures. On random directed networks we find that the critical behavior of DERP depends sensitively on degree correlations.  These analytical predictions are corroborated by extensive Monte Carlo simulations, highlighting the profound impact of directionality and correlations on long-range connectivity in complex networks.
\end{abstract}

\maketitle

% ==========================================
% SECTION I: INTRODUCTION
% ==========================================

%\section{Introduction}
Percolation theory ~\cite{dorogovtsev2008critical,araujo2014recent,bianconi2018multilayer,millan2025topology} establishes the interplay between topology and dynamics by determining the  conditions under which a network displays a giant component. Thus, percolation has wide applications in robustness~\cite{artime2021percolation_1qgfz2,buldyrev2010catastrophic},  contagion processes \cite{sun2021competition}, brain research \cite{sun2023dynamic} and quantum communication \cite{meng2021concurrence}. While classical percolation typically assumes that connectivity is strictly ensured by links to active nearest neighbors, many physical and logical processes operate over paths of extended range ~\cite{meng2025path,PhysRevLett.133.047402,kim2026shortest,PhysRevE.110.034302,PhysRevE.108.044304,cirigliano2026dynamical,hu2025unveiling}. For instance, in quantum communication networks, entanglement can be swapped between distant qubits via a sequence of quantum repeaters, provided the total path length does not exceed a coherence threshold distance $R$~\cite{huang2025quantum_gu3txz,zhang2024fast,feng2024realization,knaut2024entanglement}.  To capture these generalized connectivity rules, the concept of Extended-Range Percolation (ERP) originally formulated for lattices 
\cite{xun2021site,xun2020bond,xun2025extended} has been recently formulated for complex networks~\cite{PhysRevE.110.034302,PhysRevE.108.044304}.

Existing studies on ERP have been largely confined to undirected graphs, leaving a significant gap in our understanding of intrinsically directed systems such as quantum communication networks. It is usually assumed that in directed networks, the percolation transition is  more complex. Indeed, in directionality of the links, requires to determine the size of the In-, Out-, and Strongly Connected Giant Components (IGC, OGC, SCGC) while on undirected network percolation is described in terms of a single giant component. Contrary to this intuition, in this work we will demonstrate that,  when connectivity is defined on the basis of shortest paths, including directionality can reduce the complexity of the percolation process, as directed, unreciprocated links will not allow backtracking of the paths.

In order to show  this notable result, we propose Directed Extended-Range Percolation (DERP)  defined on a directed network in which links are not reciprocated. We provide a theoretical framework based on  a  message-passing algorithm \cite{bianconi2018multilayer,newman2023message,karrer2014percolation} to provide  an analytic solution to DERP that is exact, provided that the network is locally tree like. 
While on undirected networks evaluating connectivity within a distance $R$ requires considering backtracking of the paths connecting trusted nodes leading to a very complex message-passing algorithm \cite{PhysRevE.110.034302,PhysRevE.108.044304}, here we demonstrate the more straightforward formulation of DERP. Indeed, DERP explicitly tracks the propagation of connectivity through directed non-backtracking paths passing through intermediate relay nodes, reducing the complexity of the underlying message passing problem. This introduces a key tradeoff between DERP and ERP. Topologically, directionality requires accounting for the combined roles of the IGC, OGC, and SCGC, rather than a single giant component as in undirected ERP. Dynamically, however, enforcing directed, non-backtracking paths simplifies the message-passing algorithm, making the analytical treatment of DERP more tractable.

In DERP, relay nodes play a fundamentally different role than in standard percolation: while untrusted nodes block connectivity when $R=1$, they instead act as active relays, enabling long-range connections without altering the network structure. Our theory of DERP yields closed-form expressions for the percolation threshold and the DERP anomalous critical exponents. Notably, these results depend strongly on in–out degree correlations. For uncorrelated networks, DERP exhibits a finite threshold and mean-field exponents even with scale-free degrees. In contrast, for maximally correlated networks, the behavior depends on the degree distribution: the threshold vanishes for power-law exponents $\gamma \in (2,3]$, and anomalous critical behavior emerges for $\gamma \leq 4$.

%In this process, relay nodes play a central role  different from standard percolation. In standard directed percolation ($R=1$), unoccupied nodes strictly obstruct connectivity. In contrast, DERP transforms these nodes into active relays, enabling long-range connectivity also without changing the effective structure of the underlying network.  Our theoretical derivation enables us to find a simple closed form expression for the percolation threshold as well as for the critical indices of DERP.
%Interestingly, we observe that these results are strongly dependent on the presence of in and out degree correlations of the directed network. For uncorrelated (UC) directed networks we observe that DERP displays always a finite percolation threshold, and mean-field critical exponents also in the presence of scale-free in- and out-degree distributions. For \textit{maximally-correlated} (MC) in- and out- degree distributions, however, we observe strong dependence on the degree distribution. In this latter case,  the percolation threshold vanishes for power-law degree distributions with power-law exponent $\gamma\in (2,3]$ and we observe anomalous critical indices as long as $\gamma\leq 4$.  

{\it Directed Extended Range Percolation-} We consider a directed graph $G=(V,E)$ composed of a set $V$ of nodes and a set $E$ of directed, unreciprocated edges. 
Each node is independently assigned to be \emph{trusted} with probability $p$ or \emph{untrusted} with probability $1-p$. 
Directed extended-range percolation (DERP) studies the emergence of macroscopic connectivity between trusted nodes when communication is allowed through directed paths of maximum length $R$, possibly traversing intermediate untrusted relay nodes.
{Figure \ref{figure1} shows the mechanism underlying DERP. A trusted node $i$ connects to the network through incoming and outgoing chains of untrusted relays of lengths $m$ and $n$. 
If both $m<R$ and $n<R$, node $i$ belongs to the SCGC, while satisfying only one of these conditions places it in the IGC or OGC. %This asymmetry, absent in undirected ERP, is a key feature of directed networks and determines how the SCGC emerges from the IGC and OGC. 
%Panels (b) and (c) show that increasing the interaction range from $R=2$ to $R=3$ expands all three giant components.}

\begin{figure}[htbp]
    \centering
    \includegraphics[width=0.95\columnwidth]{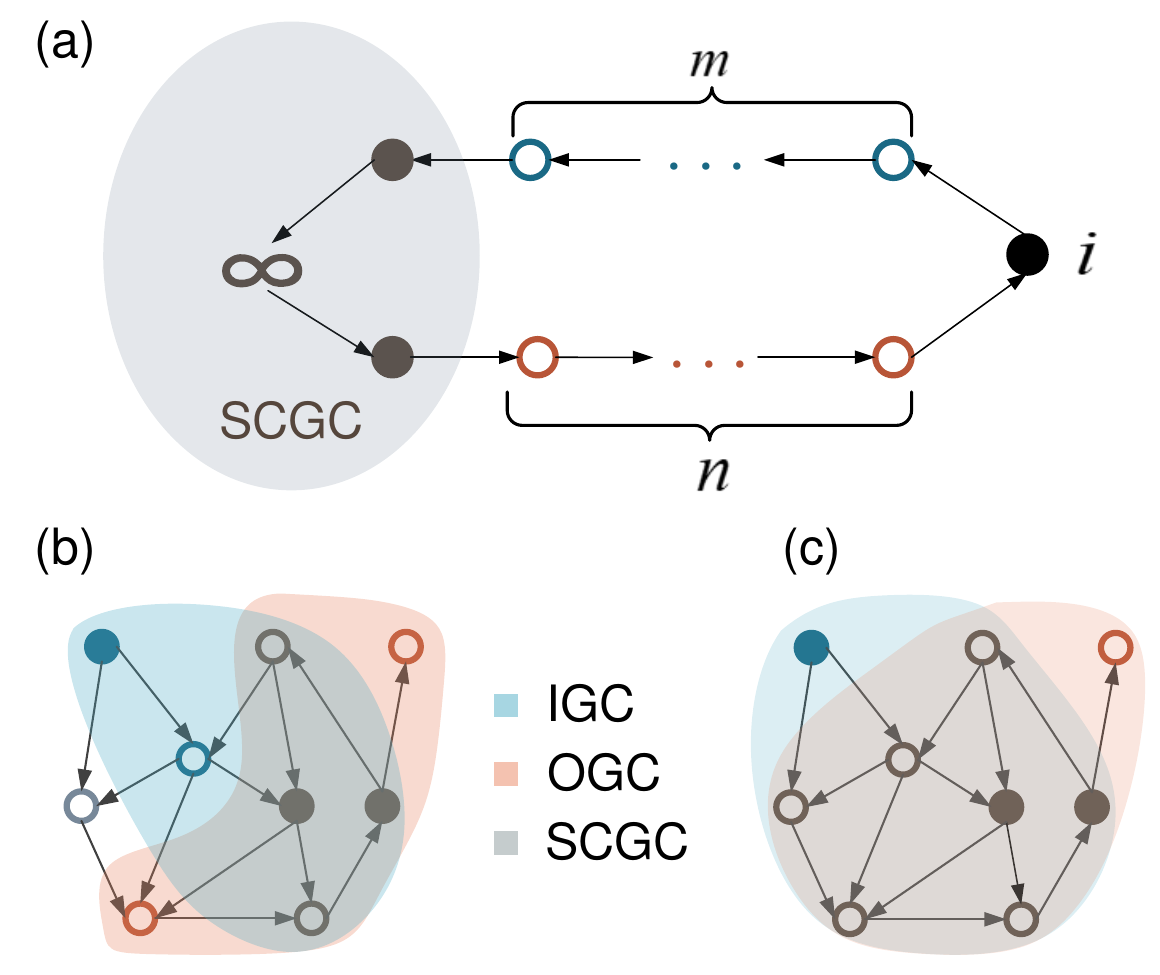}
   \caption{\label{figure1}
\textbf{DERP message passing mechanism and network components.}(a) Schematic of the DERP connectivity mechanism. Node $i$ is trusted (filled circle), while the unfilled circles are untrusted relay nodes. If both $m<R$ and $n<R$, node $i$ connects to the SCGC through incoming and outgoing paths within range $R$ and thus belongs to the SCGC together with the relay nodes. If only $m<R$ (only $n<R$), node $i$ and the corresponding relay nodes belong to the IGC (OGC). 
Spatial distribution of IGC, OGC, and SCGC in a representative network for $R=2$ (b) and $R=3$ (c), showing that increasing the interaction range expands all three giant components.}
\end{figure}

{\it Message passing approach-}
Here we formulate the message passing theory that predicts the size of the In-Giant Component, Out-Giant Components  and of the SCGC of  DERP  on a generic locally tree-like directed networks in which directed edges are not reciprocated. Due to the absence of reciprocal edges, directed paths cannot immediately backtrack, which considerably simplifies the message passing formulation compared to the undirected extended-range percolation problem \cite{PhysRevE.110.034302}. The message passing algorithm implements the following algorithm. If a node is trusted, it will connect a downstream node to the IGC if and only if the node is itself connected (by directed paths pointing to it), to trusted nodes at distance less or equal to $R$. If a node is untrusted, the node will inform a downstream node that the minimum distance to an upstream trusted node, (connected to the IGC via directed paths), is at distance $0<r<R$. These simple arguments, and their generalizations to treat OGC, lead directly to the  message passing equations. These are  recursive equations for the messages $\omega_{i\to j}^{r,(\pm)}\in [0,1]$ indicating the probability that a generic node $i$ connects   its downstream $(+)$ or upstream $(-)$ neighbor node $j$  to trusted nodes belonging to the IGC $(+)$ or OGC $(-)$  at minimum distance $r$ (with $1\leq r\leq R )$ from node $j$. 
By generalizing the algorithm valid for undirected ERP to DERP,  the  message passing equations for DERP can be derived to be
%\begin{widetext}
\bea
{\omega}_{i\to j}^{1,(\pm)}&=&p\left[1-\prod_{\ell\in N^{(\mp)}(i)\setminus j}(1-\sum_{1\leq q\leq R}{\omega}_{\ell\to i}^{q,(\pm)}))\right],\nonumber \\
{\omega}_{i\to j}^{r+1,(\pm)}&=&(1-p) \left\{\left[\prod_{\ell\in N^{(\mp)} (i)\setminus j}\left(1-\sum_{1\leq q\leq r-1}{\omega}_{\ell\to i}^{q,(\pm)}\right)\right.\right.\nonumber \\
&&\left.\left.-\prod_{\ell\in N^{(\mp)}(i)\setminus j}\left(1-\sum_{1\leq q\leq r}{\omega}_{\ell\to i}^{q,(\pm)}\right)\right]\right\},
\label{eq:MP_exact}
\eea
where  we denote by $N^{(+)}(i)$ the set of outgoing neighbors of node $i$ and by $N^{(-)}(i)$ the set of its incoming neighbors. 
This formulation explicitly delineates two distinct connectivity mechanisms: direct linkage via trusted nodes ($p$) and extended transmission through untrusted intermediates ($1-p$). Crucially, the $(1-p)$ term in Eq.~(\ref{eq:MP_exact}) represents a transmission channel rather than a vacancy. Unlike standard site percolation where unoccupied sites strictly terminate connectivity, here 'untrusted' nodes function as active relays, preserving the path continuity up to range $r+1$. This mechanism effectively enables the bridging of distant trusted clusters through an untrusted substrate, creating effective long-range correlations in the sparse network. 

The structural integrity of directed networks is characterized by the size of the SCGC. A node belongs to the SCGC if and only if it is part of both the IGC and the OGC.
Let $s_i^{(\pm)}(u_i^{(\pm)})$ denote the probability that a node $i$ is  trusted (untrusted) and belongs to the IGC ($+$) or OGC ($-$). This requires at least one neighbor to be connected to the respective component within range $R$:
\bea
\hspace{-4mm}s^{(\pm)}_i&=&p\left[1-\prod_{\ell\in N^{\mp}(i)}(1-\sum_{1\leq r\leq R}{\omega}_{\ell\to i}^{r,(\pm)})\right],
\label{eq:ri}\\
\hspace{-4mm}u^{(\pm)}_i&=&{(1-p)}\left[1-\prod_{\ell\in N^{(\mp)}(i)}(1-\sum_{1\leq r\leq R-1}{\omega}_{\ell\to i}^{r,(\pm)})\right].
\label{eq:ui}
%s_i&=&{p}\left[1-\prod_{\ell\in N^{(-)}(i)}(1-\sum_{1\leq r\leq R}{\omega}_{\ell\to i}^{r,(+)})\right]\nonumber \\
%&&\times\left[1-\prod_{\ell\in N^{(+)}(i)}(1-\sum_{1\leq r\leq R}{\omega}_{\ell\to i}^{r,(-)})\right],
%\label{eq:P}\\
%u_i&=&{(1-p)}\left[1-\prod_{\ell\in N^{(-)}(i)}(1-\sum_{1\leq r\leq R-1}{\omega}_{\ell\to i}^{r,(+)})\right]\nonumber \\
%&&\times\left[1-\prod_{\ell\in N^{(+)}(i)}(1-\sum_{1\leq r\leq R-1}{\omega}_{\ell\to i}^{r,(-)})\right].
%\label{eq:U}
%\eea
\eea
Given that a node in the SCGC by definition belongs to both the IGC and the OGC, we obtain that the probability $s_i$ that a node $i$ is trusted and belongs to the SCGC and the probability $u_i$ that a node is untrusted and belongs to the SCGC is given by
\bea
s_i = \frac{1}{p} s_i^{(+)} s_i^{(-)}, \quad u_i = \frac{1}{(1-p)}u_i^{(+)} u_i^{(-)}.%u_i = \frac{1}{(1-p)}\sum_{i} u_i^{(+)} u_i^{(-)}.
\eea
The macroscopic order parameters $S$ (fraction of trusted nodes in SCGC) and $U$ (fraction of untrusted nodes in SCGC) are finally given by 
\begin{equation}
\label{eq:SCGC_Order}
S = \frac{1}{N}\sum_{i} s_i, \quad U = \frac{1}{N}\sum_{i} u_i.
\end{equation}
\begin{figure}[htbp]
    \centering
    \includegraphics[width=0.95\columnwidth]{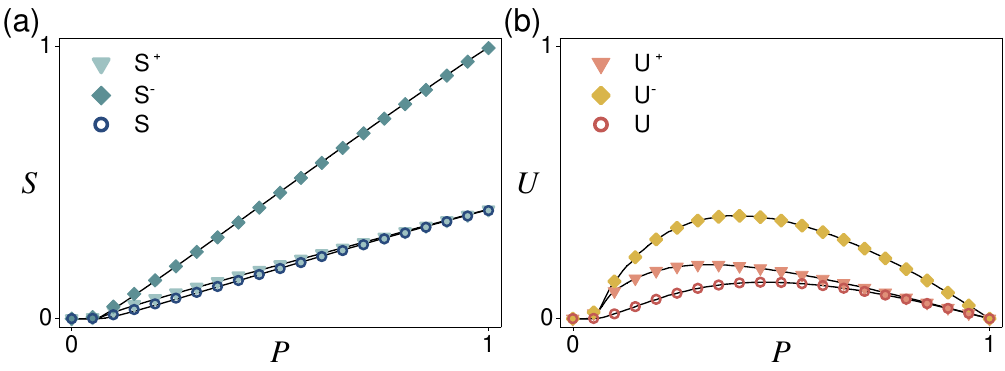}
    \caption{\label{figure2}
    \textbf{DERP dynamics in a real-world p2p network \cite{snapnets} ($N = 10,876$, $R=2$).}
    Symbols denote numerical simulations, and solid lines represent theoretical predictions. Despite the inherent network asymmetry causing the sizes of IGC and OGC to be unequal, the theoretical fits remain excellent.%\wenbo{Because real-world networks generally possess an extremely large GWCC and a small GSCC, it is easy to observe a high degree of overlap within $S$.}\gin{}
    }
    \end{figure}
DERP leads to a continuous second order phase transition well captured by the Message Passing solution as can be tested on locally tree-like real-world networks leading to excellent agreement between the simulation results and the theory (see Figure \ref{figure2}). The  critical properties of the DERP transition can be explored by
linearizing the message passing equations close to the trivial solution $S=U=0$.
The linearization of the message passing Eqs.(\ref{eq:MP_exact}) for $0<{\omega}_{i\to j}^{r,(\pm)}\ll 1$ valid for $0<p-p_c\ll1$, leads to
\bea
{\omega}_{i\to j}^{1,(\pm)}&=&p\sum_{\ell\in N^{(\mp)}(i)\setminus j}\sum_{1\leq q\leq R}{\omega}_{\ell\to i}^{q,(\pm)}\nonumber \\
{\omega}_{i\to j}^{r+1,(\pm)}&=&(1-p) \sum_{\ell\in N^{(\mp)} (i)\setminus j}{\omega}_{\ell\to i}^{r,(\pm)}.
\eea
Thus this linear system can be expressed in matrix form for the vector $\bm\omega_{i\to j}^{(\pm)}=(\omega^{1,(\pm)}_{i\to j},\omega^{2,(\pm)}_{i\to j}\ldots \omega^{R,(\pm)}_{i\to j})$ as 
\bea
\bm\omega^{(\pm)}_{i\to j}=\boldsymbol{\mathcal{A}}^{(\pm)}\bm\omega^{(\pm)}_{i\to j}
\eea
with the matrix $\mathcal{A}^{(\pm)}$ having a $R\times R$ block form of the type
\bea
\hspace{-5mm}\boldsymbol{\mathcal{A}}^{(\pm)}=\begin{pmatrix} p \mathcal{B}^{(\pm)}&p \mathcal{B}^{(\pm)}&p \mathcal{B}^{(\pm)}&p \mathcal{B}^{(\pm)}&\ldots&p \mathcal{B}^{(\pm)}&p \mathcal{B}^{(\pm)}\\
0&q \mathcal{B}^{(\pm)}&0&0&\ldots &0&0\\
0&0&q\mathcal{B}^{(\pm)}&0&\ldots &0&0\\
0&0&0&q\mathcal{B}^{(\pm)}&\ldots &0&0\\
%\vdots&\vdots&\vdots&\vdots&\vdots&\vdots&\vdots\\
\vdots&\vdots&\vdots&\vdots&\vdots&\vdots&\vdots\\
0&0&0&0&\ldots&q\mathcal{B}^{(\pm)}&0\\
0&0&0&0&\ldots&0&q\mathcal{B}^{(\pm)}
\end{pmatrix},
\eea
where $q=1-p$ and where $\mathcal{B}^{(\pm)}$ indicate the non backtracking matrices of elements
 given by 
\bea
{[\mathcal{B}^{(\pm)}]}_{(i\arrowpm j);(r\arrowpm s)}&=&p\delta_{j,r}(1-\delta_{i,s}).
\label{B}
\eea
where   $i\arrowpm j$ indicates  links going in the same $(+)$  or in the opposite $(-)$ direction of the link $(i,j)$ and here and in the following $\delta_{x,y}$ indicates the Kronecker delta.
The percolation threshold $p=p_c$ satisfies the condition $
\Lambda^{(\pm)}(p_c)=1,$ where $\Lambda^{(\pm)}$ is the maximum eigenvalue of $\boldsymbol{\mathcal{A}}^{(\pm)}$ studied as a function of $p$.
Note that given the structure of the matrices $\boldsymbol{\mathcal{A}}^{(\pm)}$ we have 
$\Lambda^{(+)}(p)=\Lambda^{(-)}(p)=\Lambda(p).$
This maximum eigenvalue can be expressed in terms of the maximum eigenvalue $\lambda$ of the non-backtracking matrix ${\mathcal{B}^{(+)}}$ which coincides with the maximum eigenvalue $\lambda$ of the non-backtracking matrix $\mathcal{B}^{(-)}$. Exploiting the block structure of the matrices $\boldsymbol{\mathcal{A}}^{(\pm)}$  we thus obtain
%\bea
%\Lambda(p)=\lambda\sum_{h=0}^{R-1}[(1-p_c)\lambda]^h=p\lambda\frac{[(1-p)\lambda]^R-1}{(1-p)\lambda-1}.
%\eea
%Thus the percolation threshold $p=p_c$ satisfies 
\bea
\Lambda(p_c)=p_c\lambda \frac{[(1-p_c)\lambda]^R-1}{(1-p_c)\lambda-1}=1.
\label{eq:p_c_MP}
\eea

{\it DERP on random networks with given degree distribution-} For  directed networks in the random network ensemble with given degree distribution $P(\mathbf{k}) = P(k_{\text{in}}, k_{\text{out}})$, the local message passing equations provide results that are self-averaging in the infinite network limit.
Let us define $W_{r}^{(\pm)}$ as the average of the message $\omega_{i\to j}^{r,(\pm)}$ over the probability distribution of the random network, i.e. $W_r^{(\pm)} =
\Avg{ \omega_{i\to j}^{r,(\pm)}}$. The ensemble average equations of the message passing Eqs. (\ref{eq:MP_exact}) valid for DERP, yields the self-consistent system of equations for $W_r^{(+)}$
\bea
W_1^{(+)}&=&p \sum_{{\bf k}}\frac{k_{\text{out}}}{\Avg{k_{\text{out}}}}
P({\bf k})\left[1-\left(1-\sum_{q=1}^{R} W_q^{(+)}\right)^{k_{\text{in}}}\right],
\nonumber \\
W_{r+1}^{(+)} &=&(1-p)\sum_{{\bf k}}\frac{k_{\text{out}}}{\Avg{k_{\text{out}}}} P({\bf k})
\left[\left(1-\sum_{q=1}^{r-1} W_q^{(+)}\right)^{k_{\text{in}}}\right.\nonumber \\
&&-
\left(1-\sum_{q=1}^{r} W_q^{(+)}\right)^{k_{\text{in}}}\Bigg],
\label{eq:Wself_plus}
\eea
while for $W_r^{(-)}$ the analogous equations hold (see SM for details). 
%\bea
%W_1^{(-)} &=&p \sum_{{\bf k}}\frac{k_{\text{in}}}{\langle k_{\text{in}}\rangle}
%P({\bf k})\left[1-\left(1-\sum_{q=1}^{R} W_q^{(-)}\right)^{k_{\text{out}}}\right],
%\nonumber \\
%W_{r+1}^{(-)} &=&(1-p)\sum_{{\bf k}}\frac{k_{\text{in}}}{\langle k_{\text{in}}\rangle} P({\bf k})
%\left[\left(1-\sum_{q=1}^{r-1} W_q^{(-)}\right)^{k_{\text{out}}}\right.\nonumber \\
%&&-
%\left(1-\sum_{q=1}^{r} W_q^{(-)}\right)^{k_{\text{out}}}\Bigg],
%\label{eq:Wself_minus}
%\eea
The fraction of nodes $S^{(+)}$ ($U^{+}$) that are trusted (untrusted) and belong to the IGC  is given by 
\bea
\hspace{-10mm}S^{(+)} &=&p \sum_{\bf k}P({\bf k})
\left[1- \left(1-\sum_{q=1}^{R} W_q^{(+)}\right)^{k_{\text{in}}}\right],\nonumber\\
\hspace{-10mm}U^{(+)}&=&(1-p) \sum_{\bf k}P({\bf k})
\left[1-\left(1-\sum_{q=1}^{R-1} W_q^{(+)}\right)^{k_{\text{in}}}\right],
\label{eq:SU_random_in}
\eea
while the analogous expression hold for the  fraction of nodes $S^{(-)}$ ($U^{-}$) that are trusted (untrusted) and belong to the OGC (see SM for details).
%which are given by 
%\bea
%\hspace{-10mm}S^{(-)}&=&p \sum_{\bf k}P({\bf k})
%\left[1-\left(1-\sum_{q=1}^{R} W_q^{(-)}}\right)^{k_{\text{out}}}\right],\nonumber\\
%\hspace{-10mm}U^{(-)}&=&(1-p) \sum_{\bf k}P({\bf k})
%\left[1-\left(1-\sum_{q=1}^{R-1} W_q^{(-)}}\right)^{k_{\text{out}}}\right].
%\label{eq:SU_random_out}
%\eea
Finally the order parameters $S$ (and $U$) indicating  the fraction of nodes that are trusted (untrusted) and belong to the SCGC is obtained from the joint probability that nodes are in the IGC and OGC, thus obtaining 
\bea
S=\frac{1}{p}S^{(+)}S^{(-)},\quad U=\frac{1}{1-p}U^{(+)}U^{(-)}.
\eea

 For random networks with a given degree distribution $P({\bf k})$, the analytical condition for criticality, determining the percolation threshold $p_c$ is obtained by linearizing the self-consistent Eqs.(\ref{eq:Wself_plus}) and their analogous equations for $W_r^{(-)}$ by taking into account that for directed network we have $\avg{k_{\text{in}}}=\avg{k_{\text{out}}}$. In this way we obtain that the percolation threshold $p_c$ satisfies
\begin{equation}
\label{eq:pc_exact}
p_c\kappa\sum_{h=0}^{R-1}[(1-p_c)\kappa]^h=p_c \kappa \frac{[(1-p_c)\kappa]^R - 1}{(1-p_c)\kappa - 1} = 1,
\end{equation}
where $\kappa$ is the branching  factor  $\kappa = {\Avg{ k_{\text{in}}k_{\text{out}}}}/{\Avg{ k_{\text{in}}}}.$
Interestingly, the expression for the percolation threshold on random directed networks given by Eq.(\ref{eq:pc_exact}) has the same structure as the expression derived in Eq.(\ref{eq:p_c_MP}) for individual locally tree-like directed networks via the message passing approach, where $\lambda$ is identified with the branching factor $\kappa$. Overall this theoretical treatment of DERP  elucidates the physical role of the interaction range $R$: it acts as an effective amplifier of the  effective connectivity. Even for sparse networks where standard percolation ($R=1$) would fail ($p_c > 1$), a finite range $R > 1$ can drive $p_c$ to much lower values, significantly increasing the network robustness.

The critical behavior of DERP is characterized by the set of critical indices indicating the scaling of the order parameters for $0<p-p_c\ll 1$:
\bea
& U^{(\pm)}\propto (p-p_c)^{\beta_U^{(\pm)}},\quad S^{(\pm)}\propto (p-p_c)^{\beta^{(\pm)}_S},\nonumber \\
& U\propto (p-p_c)^{\beta_U},\quad S\propto (p-p_c)^{\beta_S}.
\eea

{\it DERP on UC and MC Directed Networks-}
\begin{table}[htbp]
\centering
\caption{Critical exponents for UC (with identical in- and out-degree distribution) and MC directed networks for well-behaved (WB) and power-law (PL) in- and out-degree distributions. WB directed networks   have in and out degree distributions with finite  moments of order up to three and  include  Poisson directed networks.  PL directed networks have  power-law  in- and out- degree distributions  with power-law exponent $\gamma\in (2,4)$. Power-law networks with exponent $\gamma\in (2,3]$ are scale-free (SF). The critical indices are expressed in terms $\eta_r$ given by  $\eta_r = (\gamma - 2)^{r}$.}
\label{CriticalIndexTable}
\renewcommand{\arraystretch}{1.8} % 基础行距
\begin{tabular}{llcccc}
\toprule
Model & Condition & $\beta_{S}^{\pm}$ & $\beta_{U}^{\pm}$ & $\beta_{S}$ & $\beta_{U}$ \\
\midrule
\makecell[l]{UC (WB) \\ MC (WB)} & \makecell{} & 1 & 1 & 2 & 2 \\
\midrule
\multirow{2}{*}{UC (PL)} 
& $\gamma > 3$       & 1 & 1 & 2 & 2 \\[1ex] % 这里加了 1ex 间距让上下更协调
& $\gamma \in (2,3)$ & $\dfrac{1}{\gamma - 2}$ & $\dfrac{1}{\gamma - 2}$ & $\dfrac{2}{\gamma - 2}$ & $\dfrac{2}{\gamma - 2}$ \\
\midrule
\multirow{3}{*}{MC (PL)} 
& $\gamma > 4$       & 1 & 1 & 2 & 2 \\[1.5ex] % 单独增加 SFMC 内部行间距
& $\gamma \in (3,4)$ & $\dfrac{1}{\gamma - 3}$ & $\dfrac{1}{\gamma - 3}$ & $\dfrac{2}{\gamma - 3}$ & $\dfrac{2}{\gamma - 3}$ \\[2ex] % 单独增加 SFMC 内部行间距，大型分式给 2ex
& $\gamma \in (2,3)$ & $1+\dfrac{\eta_{R-1}}{1-\eta_R}$ & $\dfrac{\eta_{R-2}}{1-\eta_R}$ & $1+2\dfrac{\eta_{R-1}}{1-\eta_R}$ & $\dfrac{2\eta_{R-2}}{1-\eta_R}$ \\
\bottomrule
\end{tabular}
\vspace{1ex}
\raggedright.
\end{table}
Our theory predicts a distinct difference in the critical properties of DERP defined on uncorrelated (UC) and maximally correlated (MC) directed networks. While for UC random directed networks the in-degree and out-degree of each node are uncorrelated, the  MC random  networks impose that the in-degree of any node $i$ matches its out-degree,i.e. $k^{\text{in}}_i = k^{\text{out}}_i$. This maximal correlation changes drastically the critical properties of DERP. 

Specifically, for UC random directed networks the joint degree distribution $P({\bf k})$ factorizes into the in-degree distribution $P_{in}(k_{\text{in}})$ and the out-degree distribution $P_{out}(k_{\text{out}})$, i.e. $P({\bf k})=P_{in}(k_{\text{in}})P_{out}(k_{\text{out}})$ thus the branching ratio is simply given by $\kappa=\Avg{k_{\text{in}}}=\Avg{k_{\text{out}}}$. This implies that the critical threshold $p_c$ given by Eq.(\ref{eq:pc_exact}) is finite for any finite value of $R$, regardless of the considered in-degree and out-degree distributions $P_{in}(k_{\text{in}})$, and $P_{out}(k_{\text{out}})$. Moreover, in the case of UC random directed networks,  the critical $\beta^{(\pm)}$ and $\beta$ determining the critical behavior of the order parameters  $S^{(\pm)}$ ($U^{(\pm)}$) and $S$ ($U$) remain always mean-field (see Table \ref{CriticalIndexTable} and Supplementary Material (SM) for details of the derivation).

%\gin{It is a good idea to add a nice table here with the critical exponnet for DU and for MC}

\begin{figure}[htbp]
    \centering
    \includegraphics[width=0.95\columnwidth]{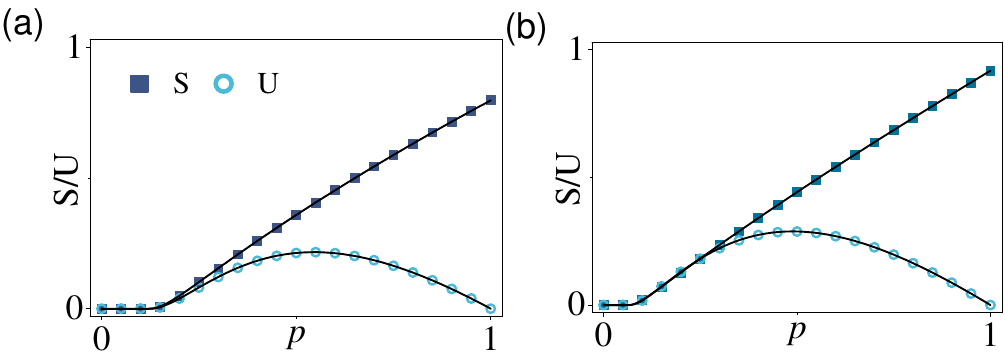}
    \caption{\label{figure3}
    \textbf{DERP dynamics in directed networks with varying degree correlations.}
    Comparison of simulations (symbols) and theoretical predictions (solid lines) for UC (Poisson) (a) and MC (Poisson) (b) networks. The order parameters $S$ and $U$ are shown as functions of $p$ for $R=2$. Network parameters are $N=20{,}000$, $c  = 2.5$. Data are averaged over 100 realizations.
    }
%Comparison of simulations (symbols) and theoretical predictions (solid lines) for UC (Poisson) (a), MC (Poisson) (b), UC (SF) (c), and MC (SF) (d) networks. The order parameters $S$ and $U$ are shown as functions of $p$ for $R=2$ and $R=3$. Network parameters are $N=20{,}000$, $c  = 2.5$ for Poisson (Poisson) networks, and $\gamma = 2.5$ for SF networks. Data are averaged over 100 realizations.}
\end{figure}

%Comparison of simulations (symbols) and theoretical predictions (solid lines) for UC (Poisson) (a), MC (Poisson) (b), UC (SF) (c), and MC (SF) (d) networks. The order parameters $S$ and $U$ are shown as functions of $p$ for $R=2$ and $R=3$. Network parameters are $N=20{,}000$, $c  = 2.5$ for Poisson (Poisson) networks, and $\gamma = 2.5$ for SF networks. Data are averaged over 100 realizations.}
%\end{figure}
\begin{figure*}[htbp]
    \centering
    \includegraphics[width=0.8\textwidth]{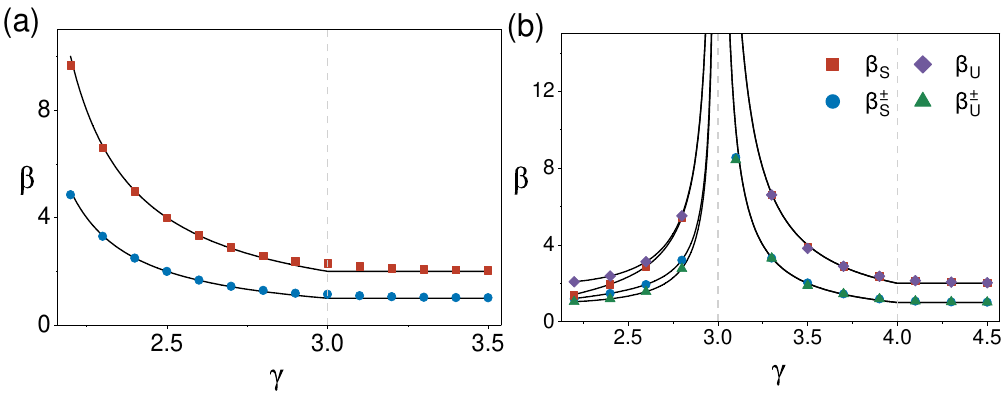}
    \caption{\label{figure4}
    \textbf{Dependence of the critical exponent $\beta$ on the parameter $\gamma$ for $R=2$ in the DERP model.} Critical exponent $\beta$ as a function of $\gamma$ for $R=2$ in the DERP model. (a) In the UC (PL) model, a single transition occurs at $\gamma=3$ (vertical dashed line). For $\gamma>3$, the system exhibits mean-field behavior. (b) In the MC (PL) model, four exponents ($\beta_S$, $\beta_U$, $\beta_S^\pm$, and $\beta_U^\pm$) show boundaries at $\gamma=3$ and $\gamma=4$, becoming non-analytic at $\gamma=3$. In both panels, symbols represent numerical simulations ($N=50,000$), and solid lines denote theoretical curves.}
\end{figure*}
On the contrary, for MC random directed networks with joint degree distribution $P({\bf k})=P(k)\delta_{k,k_{\text{out}}}\delta_{k,k_{\text{in}}}$ having branching ratio
$\kappa={\avg{k^2}}/{\avg{k}},$ we observe  that the critical behavior is strongly dependent on the degree distribution $P(k)$.  In particular, for well behaved distributions $P(k)$ having finite moments of  order up to three, we recover a finite percolation threshold $p_c$ for any finite value of $R$ and mean-field critical exponents. However, for power-law  $P(k)$ distributions with $\gamma\leq 3$ defining scale-free (SF) networks, the branching ratio  diverges in the infinite network limit, i.e. $\kappa\to \infty$  as $N\to \infty$. This implies that  the percolation threshold $p_c$ vanishes in the infinite network limit as 
\bea
p_c \simeq{\kappa^{-R}}=\left(\frac{\avg{k}}{\avg{k^2}}\right)^R\to 0.
\eea
Therefore, in DERP the convergence of the percolation threshold to zero is remarkably faster for large values of $R$ than in standard percolation on UC directed networks, recovered for $R=1$ (see SM for the perfect agreement between these predictions and our numerical simulations).
Additionally, also the critical exponents of DERP are anomalous (see Table \ref{CriticalIndexTable} and the SM  for details of the derivation) when DERP is defined on MC directed networks with power-law degree distribution $P(k)$ with  $\gamma\leq 4$ as captured in full by our theoretical predictions (see SM for details). 
To compare these results we have used advanced numerical techniques (see the Supplementary Material) for the validation of the critical indices finding excellent agreement between numerical results and analytics predictions (see Figure $\ref{figure4}$ and the SM for details).

{\it Conclusion-} Generalized path-based percolation models have attracted growing interest, yet most work focuses on undirected networks. While directionality typically increases complexity, we show that for Extended-Range Percolation (ERP), its directed counterpart (DERP) instead simplifies both the analytical treatment and the message-passing formulation, opening new directions for studying such systems.
We investigate DERP on directed networks, where connectivity between trusted nodes within distance $R$ is mediated by paths that may include untrusted intermediates. The absence of reciprocal edges suppresses backtracking, enabling a simple analytical characterization of the strongly connected giant component via message passing. As a result, DERP admits a more tractable treatment than undirected ERP, yielding exact expressions for the percolation threshold and critical exponents.
Directionality also has a strong impact on the percolation transition. The threshold decays exponentially with $R$ at large branching ratio $\kappa$, showing that even short-range extensions significantly enhance connectivity. Moreover, critical behavior depends sensitively on degree correlations. Uncorrelated networks retain mean-field behavior and a finite threshold, whereas maximally correlated scale-free networks can exhibit a vanishing threshold for $\gamma\leq 3$ and anomalous critical exponents for $\gamma\leq 4$.

%Generalized percolation model based on path connectivity are attracting increasing attention lately, however most of the work address exclusively robustness and communication in undirected networks. While in standard percolation including the directionality of the links is known to increase the complexity of the problem, here we show, somewhat counter-intuitively that for ERP, an examplar model of path-dependent percolation, its directed version, DERP simplifies the analytical solution of the model and its algorithm Message-Passing formulation. Therefore this work opens new perspective for exploring the role of directionality in these class of generalized percolation models.

%\begin{acknowledgments}
This work was supported by the National Natural Science Foundation of China under Grants No. T2422010 and No. 62172170, and by the Fundamental Research Funds for the Central Universities.
%\end{acknowledgments}
\bibliography{reference}

\newpage

\renewcommand\theequation{{S-\arabic{equation}}}
\renewcommand\thetable{{S-\Roman{table}}}
\renewcommand\thefigure{{S-\arabic{figure}}}
\setcounter{equation}{0}
\setcounter{figure}{0}
\setcounter{section}{0}

\onecolumngrid
\vspace{0.7cm}
\begin{center}
{\Large\bf             SUPPLEMENTAL MATERIAL}
\end{center}

%\appendix
\section{Critical indices}
\label{Ap}

In this section we derive the critical indices  of the DERP process on random directed networks with given degree distribution $P({\bf k})$. 

Our starting point will be the self-consistent equations for $W_r^{(+)}$ given by 
\bea
W_1^{(+)}&=&p \sum_{{\bf k}}\frac{k_{\text{out}}}{\Avg{k_{\text{out}}}}
P({\bf k})\left[1-\left(1-\sum_{q=1}^{R} W_q^{(+)}\right)^{k_{\text{in}}}\right],
\nonumber \\
W_{r+1}^{(+)} &=&(1-p)\sum_{{\bf k}}\frac{k_{\text{out}}}{\Avg{k_{\text{out}}}} P({\bf k})
\left[\left(1-\sum_{q=1}^{r-1} W_q^{(+)}\right)^{k_{\text{in}}}-
\left(1-\sum_{q=1}^{r} W_q^{(+)}\right)^{k_{\text{in}}}\right],
\label{eq:Wself_plus_A}
\eea
and the analogous one for $W_r^{(-)}$ given by 
\bea
W_1^{(-)} &=&p \sum_{{\bf k}}\frac{k_{\text{in}}}{\langle k_{\text{in}}\rangle}
P({\bf k})\left[1-\left(1-\sum_{q=1}^{R} W_q^{(-)}\right)^{k_{\text{out}}}\right],
\nonumber \\
W_{r+1}^{(-)} &=&(1-p)\sum_{{\bf k}}\frac{k_{\text{in}}}{\langle k_{\text{in}}\rangle} P({\bf k})
\left[\left(1-\sum_{q=1}^{r-1} W_q^{(-)}\right)^{k_{\text{out}}}-
\left(1-\sum_{q=1}^{r} W_q^{(-)}\right)^{k_{\text{out}}}\right],
\label{eq:Wself_minus_A}
\eea
The order parameters $S^{(+)}$ ($U^{+}$) indicating the fraction of nodes  that are trusted (untrusted) and belong to the IGC  are given by 
\bea
\hspace{-10mm}S^{(+)} &=&p \sum_{\bf k}P({\bf k})
\left[1- \left(1-\sum_{q=1}^{R} W_q^{(+)}\right)^{k_{\text{in}}}\right],\nonumber\\
\hspace{-10mm}U^{(+)}&=&(1-p) \sum_{\bf k}P({\bf k})
\left[1-\left(1-\sum_{q=1}^{R-1} W_q^{(+)}\right)^{k_{\text{in}}}\right],
\label{eq:SU_random_in_A}
\eea
The analogous expression hold for the  fraction of nodes $S^{(-)}$ ($U^{-}$) that are trusted (untrusted) and belong to the OGC which are given by 
\bea
\hspace{-10mm}S^{(-)}&=&p \sum_{\bf k}P({\bf k})
\left[1-\left(1-\sum_{q=1}^{R} W_q^{(-)}\right)^{k_{\text{out}}}\right],\nonumber\\
\hspace{-10mm}U^{(-)}&=&(1-p) \sum_{\bf k}P({\bf k})
\left[1-\left(1-\sum_{q=1}^{R-1} W_q^{(-)}\right)^{k_{\text{out}}}\right].
\label{eq:SU_random_out_A}
\eea
Finally, the order parameters $S$ (and $U$) indicating  the fraction of nodes that are trusted (untrusted) and belong to the SCGC is obtained from the joint probability that nodes are in the IGC and OGC, thus obtaining 
\bea
S=\frac{1}{p}S^{(+)}S^{(-)},\quad U=\frac{1}{1-p}U^{(+)}U^{(-)}.
\eea
In order to study the critical behavior of DERP described by these set of equations,  let us distinguish between different scenarios depending on the moments of $P({\bf k})$.
\subsection{Well behaved degree distributions}
\label{wb}
If the degree distribution $P({\bf k})$ is well behaved meaning that all its moments are finite, we can expand the self-consistent Eqs.(\ref{eq:Wself_plus_A}) and  (\ref{eq:Wself_minus_A}) close to the percolation threshold, i.e. for $0<p-p_c\ll1 $ when $0<W_{r}^{(\pm)}\ll 1$ obtaining,

%\begin{widetext}
\bea
{W}_{1}^{(+)}&\simeq&p\left[\kappa\left(\sum_{1\leq q\leq R}W_{q}^{(+)}\right)-\frac{1}{2}\kappa_{2,in}\left(\sum_{1\leq q\leq R}W_{q}^{(+)}\right)^2\right],\nonumber \\
W_{r+1}^{(+)}&\simeq&(1-p)\kappa W_{r}^{(+)},\nonumber \\
{W}_{1}^{(-)}&\simeq&p\left[\kappa\left(\sum_{1\leq q\leq R}W_{q}^{(-)}\right)-\frac{1}{2}\kappa_{2,out}\left(\sum_{1\leq q\leq R}W_{q}^{(-)}\right)^2\right],\nonumber \\
W_{r+1}^{(-)}&\simeq&(1-p)\kappa W_{r}^{(-)},\nonumber \\
%{W}_{1}^{(-)}&=&p\sum_{\bf k}\frac{k^{in}}{\avg{k^{in}}}P({\bf k})\left[1-\left(1-\sum_{1\leq q\leq R}W_{q}^{(-)}\right)^{k^{out}}\right],\nonumber \\
%W_{r+1}^{(-)}&=&(1-p)\sum_{\bf k}\frac{k^{in}}{\avg{k^{in}}}P({\bf k})\left[\left(1-\sum_{1\leq q\leq r-1}W_{q}^{(-)}\right)^{k^{out}}-\left(1-\sum_{1\leq q\leq r}W_{q}^{(-)}\right)^{k^{out}}\right].
\label{mes_ensemble_linearized}
\eea
with 
\bea
\kappa_{2,in}=\frac{\Avg{k^{out}k^{in}(k^{in}-1)}}{\avg{k^{out}}},\quad \kappa_{2,out}=\frac{\Avg{k^{in}k^{out}(k^{out}-1)}}{\avg{k^{out}}}
\eea
%\end{widetext}
It follows that 
\bea
\sum_{1\leq q\leq R}W_{q}^{(\pm)}=KW_{1}^{(\pm)}\label{Kpc}
\eea
with $K$ given by
\bea
K=\frac{[(1-p)\kappa]^R-1}{(1-p)\kappa-1}.
\eea
Moreover we have 
\bea
W_{r}^{(\pm)}=((1-p)\kappa)^{r-1} W_{1}^{(\pm)}.\label{Recursive}
\eea
It follows that  the percolation threshold $p_c$ obeys \bea
p_c\kappa K=1\eea, i.e $p_c=1/(\kappa K)$ and that for $0<p-p_c\ll1$ we obtain
\bea
W_{r}^{(\pm)}\propto (p-p_c)^{\beta}
\eea
with 
\bea
\beta=1
\eea
for every value of $1\leq r\leq R$.
Finally also the order parameters $S^{(\pm)}, U^{(\pm)}, S, U$ will  scale as 
\bea
& S^{(\pm)}\propto(p-p_c)^{\beta_S^{(\pm)}},\quad U^{(\pm)}\propto (p-p_c)^{\beta_U^{(\pm)}},\nonumber \\
& S\propto(p-p_c)^{\beta_S},\quad U\propto (p-p_c)^{\beta_U}
\eea
with the mean field exponents:
\bea
&\beta_S^{(\pm)}=\beta_U^{(\pm)}=\beta=1\nonumber \\
&\beta_S=\beta_U=2\beta=2.
\eea

\subsection{UC Power-law directed networks}
For investigating anomalous critical behavior we consider here the case of UC Power-law directed networks
\begin{equation}
P({\bf k})=P_{in}(k_{\text{in}})P_{out}(k_{\text{out}}),\\
\end{equation}
\\
with $P_{in}(k_{\text{in}})=Ck_{\text{in}}^{-\gamma}$, $P_{out}(k_{\text{out}})=Ck_{\text{out}}^{-\gamma},$
where $C$ is a normalization constant and the  power-law exponent is greater than two, i.e. $\gamma>2$ to ensure a finite average in-degree and finite average out-degree of the network in the large network limit.

We observe that in this scenario, $\kappa$ remains always finite while $\kappa_{2,\in}$ and $\kappa_{2,out}$ are finite for $\gamma>3$ while they diverge for $\gamma\in (2,3)$. 
Therefore in the case $\gamma>3$ we recover the mean-field critical exponent of well behaved distribution discussed in Sec. \ref{wb}. For $\gamma\in (2,3)$  we consider the asymptotic expansion of self-consistent Eqs.(\ref{eq:Wself_plus_A})-(\ref{eq:Wself_minus_A}) close to the percolation threshold, i.e. for $0<p-p_c\ll1 $ when $0<W_{r}^{(\pm)}\ll 1$ obtaining for ${W}_{r,(+)}$
%\begin{widetext}
\bea
{W}_{1}^{(+)}&\simeq&p\left[\kappa\left(\sum_{1\leq q\leq R}W_{q}^{(+)}\right)-a_+\left(\sum_{1\leq q\leq R}W_{q}^{(+)}\right)^{\gamma-1}\right],\nonumber \\
W_{r+1}^{(+)}&\simeq&(1-p)\kappa W_{r}^{(+)},\eea
while for ${W}_{r,(-)}$ we obtain the analogous expansions
\bea
{W}_{1}^{(-)}&\simeq&p\left[\kappa\left(\sum_{1\leq q\leq R}W_{q}^{(-)}\right)-a_-\left(\sum_{1\leq q\leq R}W_{q}^{(-)}\right)^{\gamma-1}\right],\nonumber \\
W_{r+1}^{(-)}&\simeq&(1-p)\kappa W_{r}^{(-)},
\label{mes_ensemble_linearized}
\eea
%\end{widetext}
where $a_{\pm}$ are constants.
In this scenario Eq.(\ref{Kpc}) remains valid as well as Eq.(\ref{Recursive}) and the expression of the percolation threshold $p_c=1/(\kappa K)$.
It follows that  for $0<p-p_c\ll1 $ we observe the critical scaling
\bea
W_{r}^{(\pm)}\propto (p-p_c)^{\beta}
\eea
with 
\bea
\beta=1/(\gamma-2).
\eea
This exponent determines the scaling  for all the other order parameters as  well, i.e.
\bea
& S^{(\pm)}\propto(p-p_c)^{\beta_S^{(\pm)}},\quad U^{(\pm)}\propto (p-p_c)^{\beta_U^{(\pm)}},\nonumber \\
& S\propto(p-p_c)^{\beta_S},\quad U\propto (p-p_c)^{\beta_U}
\eea
with the  exponents:
\bea
&\beta_S^{(\pm)}=\beta_U^{(\pm)}=\beta=1/{(\gamma-2)}\nonumber \\
&\beta_S=\beta_U=2\beta={2}/{(\gamma-2)}.
\eea

Note that here we omit to discuss the case $\gamma=3$ where one expects logarithmic corrections.
\subsection{Case of MC power-law directed networks}
For investigating anomalous critical behavior we consider here the case of MC power-law directed networks
\bea
P({\bf k})=\delta_{k_{\text{in}},k}\delta_{k_{\text{out}},k}P(k)
\eea
with \bea P(k)=Ck^{-\gamma},\eea
where $C$ is a normalization constant and the  power-law exponent is greater than two, i.e. $\gamma>2$ to ensure a finite average in-degree and finite average out-degree of the network in the large network limit.
We observe that in this scenario, for $\gamma>4$ all relevant moments, $\kappa$ $\kappa_{2,\in}$ and $\kappa_{2,out}$ are finite, for $\gamma\in (3,4)$ $\kappa_{2,\in}$ and $\kappa_{2,out}$ diverges while for $\gamma\in (2,3)$ also $\kappa$ diverges.
In the  case $\gamma>4$ we recover the mean-field exponents of well behaved distributions discussed in Sec. \ref{wb} so in the following we will discuss exclusively the cases $\gamma\in (3,4)$ and $\gamma \in (2,3)$. Note that we also omit here the case $\gamma=3$ and $\gamma=4$ noting that in those cases we should observe logarithmic corrections to the critical scaling.
For $\gamma\in (3,4)$  we consider the asymptotic expansion of  the self-consistent Eqs. (\ref{eq:Wself_minus_A})-(\ref{eq:Wself_plus_A}) close to the percolation threshold, i.e. for $0<p-p_c\ll1 $ when $0<W_{r}^{(\pm)}\ll 1$ obtaining,

%\begin{widetext}
\bea
{W}_{1}^{(+)}&\simeq&p\left[\kappa\left(\sum_{1\leq q\leq R}W_{q}^{(+)}\right)-a_+\left(\sum_{1\leq q\leq R}W_{q}^{(+)}\right)^{\gamma-2}\right],\nonumber \\
W_{r+1}^{(+)}&\simeq&(1-p)\kappa W_{r}^{(+)},\nonumber \\
{W}_{1}^{(-)}&\simeq&p\left[\kappa\left(\sum_{1\leq q\leq R}W_{q}^{(-)}\right)-a_-\left(\sum_{1\leq q\leq R}W_{q}^{(-)}\right)^{\gamma-2}\right],\nonumber \\
W_{r+1}^{(-)}&\simeq&(1-p)\kappa W_{r}^{(-)},\nonumber \\
%{W}_{1}^{(-)}&=&p\sum_{\bf k}\frac{k^{in}}{\avg{k^{in}}}P({\bf k})\left[1-\left(1-\sum_{1\leq q\leq R}W_{q}^{(-)}\right)^{k^{out}}\right],\nonumber \\
%W_{r+1}^{(-)}&=&(1-p)\sum_{\bf k}\frac{k^{in}}{\avg{k^{in}}}P({\bf k})\left[\left(1-\sum_{1\leq q\leq r-1}W_{q}^{(-)}\right)^{k^{out}}-\left(1-\sum_{1\leq q\leq r}W_{q}^{(-)}\right)^{k^{out}}\right].
\label{mes_ensemble_linearized}
\eea
%\end{widetext}
where $a_{\pm}>0$ are constants.
In this scenario Eq.(\ref{Kpc}) remains valid as well as Eq.(\ref{Recursive}) and the expression of the percolation threshold $p_c=1/(\kappa K)$.
It follows that  for $0<p-p_c\ll1 $ we observe the critical scaling
\bea
W_{r}^{(\pm)}\propto (p-p_c)^{\beta}
\eea
with 
\bea
\beta=1/(\gamma-3).
\eea
This exponent is the exponents for all order parameters as  well, i.e. 
\bea
S^{(\pm)}\propto(p-p_c)^{\beta_S^{(\pm)}},\quad U^{(\pm)}\propto (p-p_c)^{\beta_U^{(\pm)}},\nonumber \\
S\propto(p-p_c)^{\beta_S},\quad U\propto (p-p_c)^{\beta_U}
\eea
with the  exponents:
\bea
&\beta_S^{(\pm)}=\beta_U^{(\pm)}=\beta={1}/{(\gamma-3)}\nonumber \\
&\beta_S=\beta_U=2\beta={2}/{(\gamma-3)}.
\eea

For $\gamma\in (2,3)$ we observe that also $\kappa$ diverges, and thus the percolation threshold vanishes $p_c=0$. Let us consider here the critical indices of the directed ERP process in this very anomalous case.
For $\gamma\in (2,3)$  we consider the asymptotic expansion of   the Eqs.(\ref{eq:Wself_minus_A}) and (\ref{eq:Wself_plus_A}) close to the percolation threshold, i.e. for $0<p-p_c\ll1 $ when $0<W_{r}^{(\pm)}\ll 1$ obtaining,

%\begin{widetext}
\bea
{W}_{1}^{(+)}&\simeq&p b\left(\sum_{1\leq q\leq R}W_{q}^{(+)}\right)^{\gamma-2},\nonumber \\
W_{r+1}^{(+)}&\simeq&(1-p)b\left[\left(\sum_{1\leq q\leq r}W_{q}^{(+)}\right)^{\gamma-2}-\left(\sum_{1\leq q\leq r-1}W_{q}^{(+)}\right)^{\gamma-2}\right],\nonumber \\
{W}_{1}^{(-)}&\simeq&pb\left(\sum_{1\leq q\leq R}W_{q}^{(-)}\right)^{\gamma-2},\nonumber \\
W_{r+1}^{(-)}&\simeq&(1-p)b\left[\left(\sum_{1\leq q\leq r}W_{q}^{(-)}\right)^{\gamma-2}-\left(\sum_{1\leq q\leq r-1}W_{q}^{(-)}\right)^{\gamma-2}\right],\nonumber \\
%{W}_{1}^{(-)}&=&p\sum_{\bf k}\frac{k^{in}}{\avg{k^{in}}}P({\bf k})\left[1-\left(1-\sum_{1\leq q\leq R}W_{q}^{(-)}\right)^{k^{out}}\right],\nonumber \\
%W_{r+1}^{(-)}&=&(1-p)\sum_{\bf k}\frac{k^{in}}{\avg{k^{in}}}P({\bf k})\left[\left(1-\sum_{1\leq q\leq r-1}W_{q}^{(-)}\right)^{k^{out}}-\left(1-\sum_{1\leq q\leq r}W_{q}^{(-)}\right)^{k^{out}}\right].
\label{mes_ensemble_linearized_g}
\eea
%\end{widetext}
where $b>0$ is constant.
It follows that 
\bea
V_{r(\pm)}=\sum_{1\leq q\leq r}W_{q}^{(\pm)}
\eea
obeys the recursive equation
\bea
V_{r+1(\pm)}=W_{1}^{(\pm)}+(1-p)b [V_{r,(\pm)}]^{\gamma-2}
\eea
with 
\bea
V_{1,(\pm)}=W_{1}^{(\pm)}.
\eea
For $\gamma\in (2,3)$ and $0<W_{1}^{(\pm)}\ll1$ the leading term is therefore 
\bea
V_{r,(\pm)}\simeq((1-p)b)^{r-1}[W_{1}^{(\pm)}]^{\eta_{r-1}}
\eea
with 
\bea
\eta_{r}=(\gamma-2)^{r}.
\eea
Thus considering the first and the third of Eqs.(\ref{mes_ensemble_linearized_g}) we get
\bea
W_{1}^{(\pm)}=pV_{R,(\pm)}^{\gamma-2}\simeq p[(1-p)b]^{(R-1)(\gamma-2)}[W_{1}^{(\pm)}]^{\eta_{R}}
\eea
which leads to the critical scaling 
\bea
W_{1}^{(\pm)}\propto p^{\beta}
\eea
with 
\bea
\beta=1/(1-\eta_{R})=\left[{1-(\gamma-2)^{R}}\right]^{-1}.
\eea
and consequently 
\bea
V_{r,(\pm)}\simeq p^{\eta{r-1}\beta}.
\eea
From this analysis it follows that  for $0<p-p_c\ll1$, the order parameters will scale as 
\bea
S^{(\pm)}\propto pV_R,\quad U^{(\pm)}\propto V_{R-1},\nonumber \\
S\propto p V_R^2,\quad U\propto V_{R-1}^2.
\eea
Thus observe the critical scaling of the order parameters with exponents

Inserting this scaling in the order parameters we get the critical indices
\bea
&S^{(\pm)}\propto p^{\beta_S^{(\pm)}},\quad U^{(\pm)}\propto p^{\beta_U^{(\pm)}}\nonumber \\
&S\propto p^{{\beta}_S},\quad U\propto p^{\beta_U} 
\eea
with 
\bea
\hspace{-10mm}\beta_S^{(\pm)}=1+\frac{\eta_{R-1}}{1-\eta_{R}}\quad \beta_U^{(\pm)}=\frac{\eta_{R-2}}{1-\eta_R},\nonumber \\
{\beta}_S=1+2\frac{\eta_{R-1}}{1-\eta_R},\quad \beta_U=2\frac{\eta_{R-2}}{1-\eta_R}.
\eea
\section{Numerical validation of the critical indices}
Here we provide numerical validation of our predicted critical indices.

{
To accurately extract the true critical exponents from our numerical results, we employed a local effective exponent analysis. In the immediate vicinity of the critical point $p_c$, the order parameter (e.g., the fraction of nodes in the giant component, $S$) obeys the power-law scaling relation $S \propto (p - p_c)^\beta$. We compute the derivative with respect to the distance to criticality, $\Delta p = p - p_c$, on a double-logarithmic scale, defining the local effective exponent $\beta_{eff}$ as:
\begin{equation}
\beta_{eff}(\Delta p) = \frac{d \ln S}{d \ln \Delta p}
\end{equation}

In our discrete numerical calculations, we selected a sequence of distances to criticality $\Delta p_i \in [10^{-5}, 10^{-1}]$ spaced evenly on a logarithmic scale, and estimated the local derivative using the finite difference method:
\begin{equation}
\beta_{eff}(\Delta p_i) \approx \frac{\log_{10} S(\Delta p_{i+1}) - \log_{10} S(\Delta p_i)}{\log_{10} \Delta p_{i+1} - \log_{10} \Delta p_i}
\end{equation}

%\textbf{Identification of the Plateau Region and Theoretical Validation:}
In numerical analysis, $\beta_{eff}$ is typically subject to the competing influences of two factors: finite-size effects coupled with numerical truncation errors when extremely close to the critical point (small $\Delta p$), and higher-order nonlinear corrections when moving further away from the critical point (large $\Delta p$). Consequently, the true critical scaling behavior only manifests within an intermediate asymptotic regime where $\beta_{eff}$ exhibits a relatively flat plateau ($d\beta_{eff}/d(\ln \Delta p) \approx 0$). By extracting the stable value within this plateau region, we effectively isolate the simulated critical exponent from both boundary interferences.

Figure \ref{figure1S} demonstrates the application of this method to both UC and MC networks with Power-Law (PL) degree distributions, yielding validation results highly consistent with our analytical theory. For UC (PL) networks (Fig. \ref{figure1S}a for $\gamma=2.5$ and Fig. \ref{figure1S}b for $\gamma=3.5$), a behavioral transition occurs at $\gamma=3$. When $\gamma>3$, the system exhibits mean-field characteristics, whereas for $2<\gamma\le3$, $\beta$ becomes explicitly dependent on $\gamma$ (as derived in Eq. S-22). 

Conversely, for MC (PL) networks (Fig. \ref{figure1S}c–e corresponding to $\gamma=2.5, 3.5, 4.5$, respectively), the critical exponents depend on both $\gamma$ and $R$ when $\gamma \in (2,3)$ (Eq. S-42). Notably, because $p_c=0$ in this specific regime (Fig. \ref{figure1S}c), the left-side deviations caused by finite-size effects are absent. Broadly, the scaling behaviors in MC (PL) for $\gamma \in (3,4)$ and $\gamma>4$ structurally mirror those of UC (PL) for $\gamma \in (2,3)$ and $\gamma>3$, respectively.
}

\begin{figure}[!htb!]
    \centering
    \includegraphics[width=0.95\columnwidth]{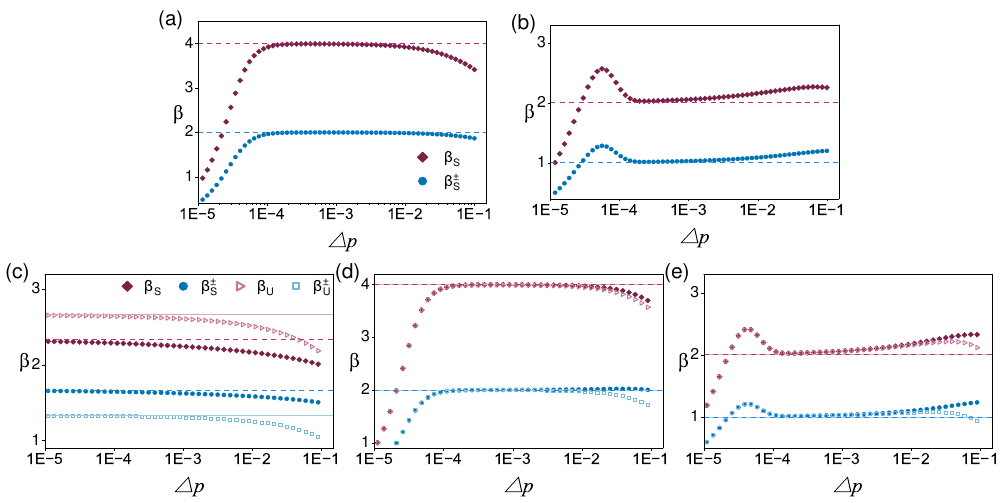}
    \caption{\textbf{Extraction of the critical exponent $\beta$ using the local effective exponent method in UC (PL) and MC (PL) networks.} The plateau regions indicate the stable simulated exponents, effectively isolating finite-size effects (left) and higher-order corrections (right). Panels (a) and (b) show the UC (PL) results for $\gamma=2.5$ and $\gamma=3.5$. Panels (c), (d), and (e) display the MC (PL) results for $\gamma=2.5, 3.5$, and $4.5$, respectively. Horizontal lines denote the analytically predicted exponents.}
    \label{figure1S}
\end{figure}

Specifically Figure \ref{figure2S} shows the order parameters as functions of $0<p-p_c\ll1$ on a log-log scale. For uncorrelated (UC) Poisson networks and for maximally correlated (MC) Poisson networks (panels a, b), the critical exponents $\{\beta_S^{(\pm)}, \beta_S, \beta_U^{(\pm)}, \beta_U\}$ with the the numerical error, match the mean-field values $\{1,2,1,2\}$ obtained from our theoretical predictions and listed in Table~I. Also for  UC (SF) networks with $\gamma=2.5 \in (2,3)$ (panel c), the exponents  agree within the numerical error with the theoretical predictions $\{1/(\gamma-2),2/(\gamma-2),1/(\gamma-2),2/(\gamma-2)\}=\{2,4,2,4\}$  For MC (SF) networks (panel d), the exponents differ from both the mean-field and UC cases, consistent with the scaling for $\gamma \in (2,3)$ in Table~\ref{CriticalIndexTable}.

\begin{figure}[!htb!]
    \centering
    \includegraphics[width=0.8\columnwidth]{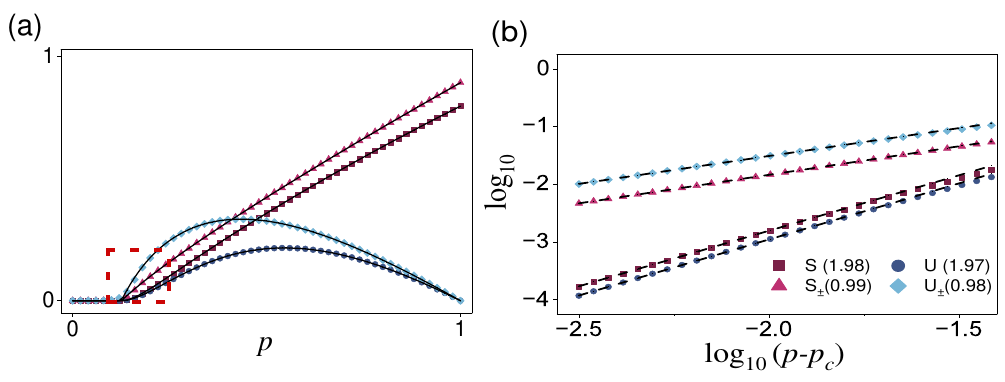}
    \caption{\textbf{Numerical evaluation of critical exponents.} (a) The order parameters $S_{\pm}$, $S$, $U_{\pm}$, and $U$ as functions of $p$ during the DERP process for a UC (WB) network with system size $N=20000$, $c=2.5$, and $R=2$. The red dashed rectangle indicates the critical region near the percolation threshold $p_c$. (b) Log-log plot of the order parameters versus $p-p_c$ ($0 < p-p_c \ll 1$), corresponding to the critical region highlighted in (a). Dashed lines represent linear fits to the data, allowing for the direct extraction of critical exponents according to definitions. The fitted exponent values are shown in parentheses in the legend.}
    \label{figure2S}
\end{figure}

\section{Further comparison between  DERP simulations and the theoretical predictions}
In this section we provide further evidence of the excellent agreement between our theoretical predictions and the Monte Carlo simulations of DERP on random UC and MC directed networks.
In Figure \ref{figure3S}  we compare the theoretical results with the Monte Carlo simulations of DERP for both UC and MC Poisson directed networks for different values of the interaction range $R$. 
In Figure \ref{figure5S} we further confirm the excellent agreement between theory and simulation results for both UC and MC scale-free directed networks for different values of the interaction range $R$.
The prediction of the percolation threshold is  also validated by extensive numerical simulations In Figure \ref{figure4S} we plot the percolation threshold $p_c$ as a function of the average degree $c$ for UC (Poisson) and MC (Poisson) networks at $R=2$.  As the network size increases we observe convergence of the percolation threshold toward the predicted random graph results Eq(\ref{figure4S}). For any given value of $c$, MC networks have lower $p_c$ than UC networks, indicating the effect of degree correlations on connectivity.

\begin{figure}[!htb!]
    \centering
    \includegraphics[width=0.7\columnwidth]{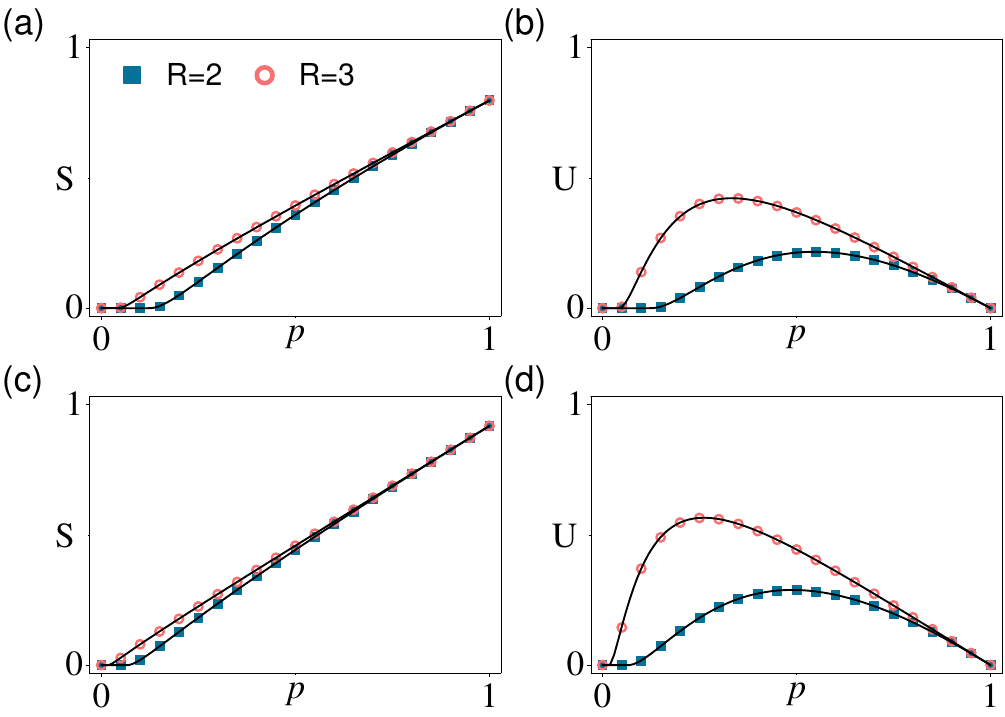}
    \caption{\textbf{DERP dynamics in directed Poisson networks with varying $R$.} Comparison of simulations (symbols) and theoretical predictions (solid lines) for UC  (a, b) and MC  (c, d) directed Poisson networks with different value of the interaction range $R$. The order parameters $S$ and $U$ are shown as functions of $p$ for $R=2$ and $R=3$. Network parameters are $N=20{,}000$ and $c = 2.5$. Data are averaged over 100 realizations.}
    \label{figure3S}
\end{figure}
\begin{figure}[htbp]
    \centering
    \includegraphics[width=0.5\columnwidth]{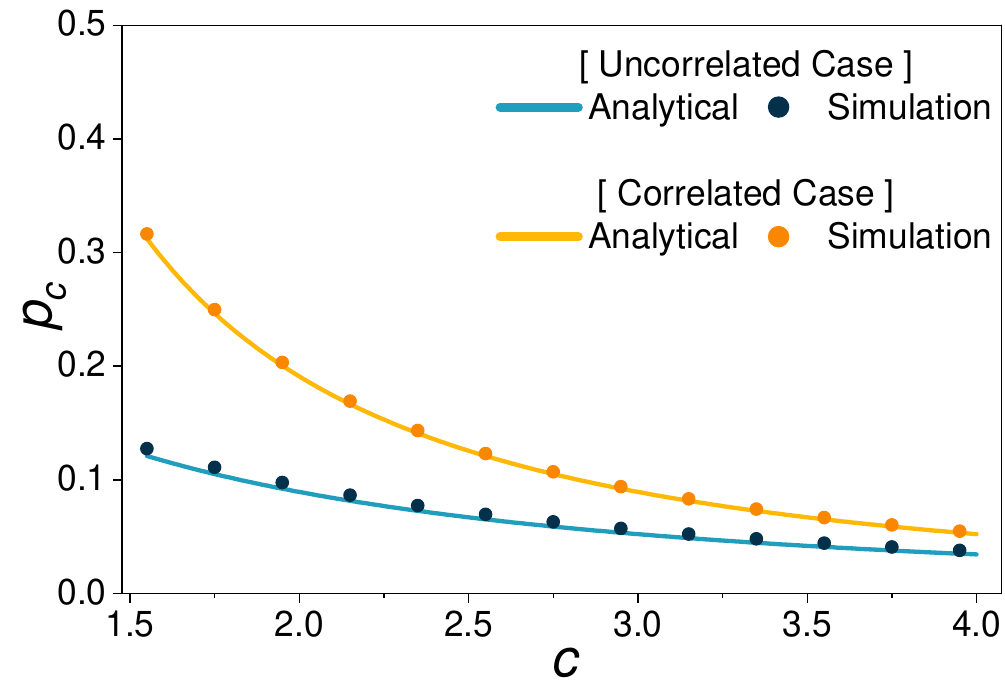}
    \caption{\label{figure4S}
    \textbf{Analytical and simulated critical thresholds.}%\gin{It would be important here to show the average of the MP predictions for single networks given by Eq.(10), this should be a perfect fit to the numerical results for any network size I believe.}
Critical threshold $p_c$ as a function of the average degree $\langle c \rangle$ for UC (Poisson) and MC (Poisson) networks ($R=2$). Symbols denote simulations for  $N=20,000$, and solid lines follow Eq.~(\ref{eq:pc_exact}). Data are averaged over 100 realizations.}
\end{figure}
\begin{figure}[!htb!]
    \centering
    \includegraphics[width=0.7\columnwidth]{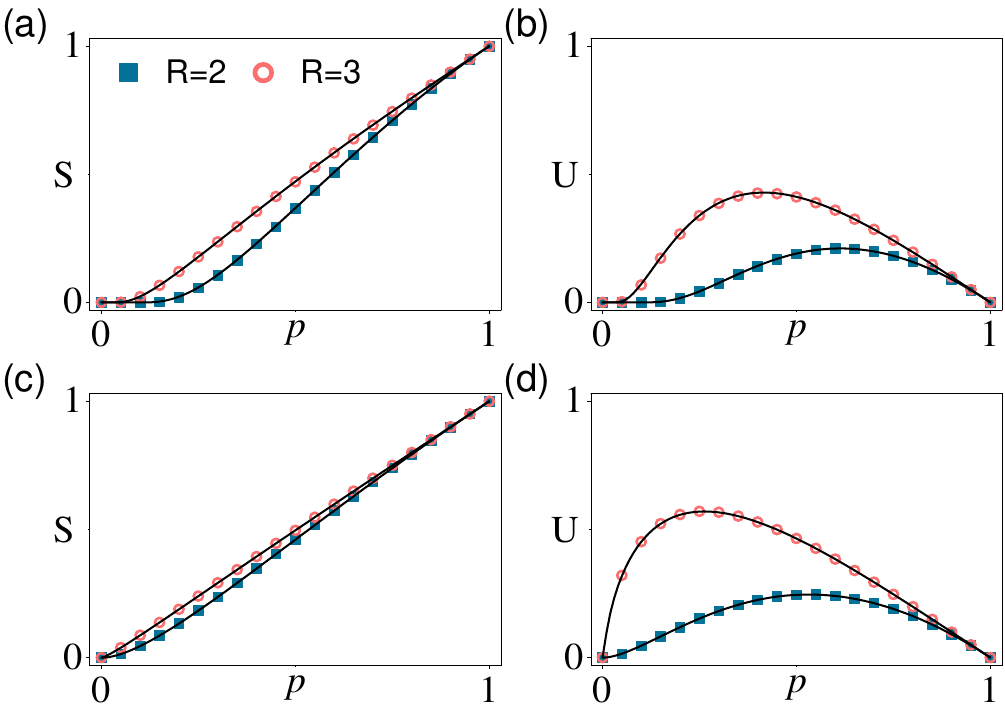}
    \caption{\textbf{DERP dynamics in directed scale-free networks with varying $R$.} Comparison of simulations (symbols) and theoretical predictions (solid lines) for UC  (a, b) and MC  (c, d) scale-free networks with different values of the interaction range $R$. The order parameters $S$ and $U$ are shown as functions of $p$ for $R=2$ and $R=3$. Network parameters are $N=20{,}000$ and the power-law exponent of the degree distribution is $\gamma = 2.5$. Data are averaged over 100 realizations.}
    \label{figure5S}
\end{figure}
\end{document}